\documentstyle[12pt]{article}
\catcode `\@=11
\@addtoreset{equation}{section}

\def\qed{\vrule height 6pt width 6pt depth 6pt}

\newcommand{\be}{\begin{equation}}
\newcommand{\ee}{\end{equation}}
\newcommand{\R}{\rm I \mkern -3mu R}
\newcommand{\N}{\rm I \mkern -3mu N}
\newtheorem{thm}{Theorem}
\newtheorem{rem}{Remark}
\newtheorem{lemma}{Lemma}
\newtheorem{cor}{Corollary}
\newtheorem{prop}{Proposition}
\begin{document}
\begin{center}
{\bf \Large{VARIATIONAL EQUATIONS AND\\
~\\SYMMETRIES IN THE LAGRANGIAN\\~\\FORMALISM II. ARBITRARY VECTOR FIELDS}}
\end{center}
\vskip 1.0truecm
\centerline{D. R. Grigore\footnote{Permanent address: Dept. Theor. Phys.,
Inst. Atomic Phys., Bucharest- M\u agurele, P. O. Box MG 6, ROM\^ANIA}
\footnote{e-mail: grigore@theor1.ifa.ro, grigore@roifa.ifa.ro}}
\vskip5mm
\centerline{Dept. of Math., Silesian Univ. at Opava}
\centerline{Bezru\u covo n\`am. 13, 74601 Opava, Czech Rep.}
\vskip 2cm
\bigskip \nopagebreak
\begin{abstract}
\noindent
We continue the study of symmetries in the Lagrangian formalism of arbitrary
order  with the help of the so-called Anderson-Duchamp-Krupka equations.
For the case of second-order equations and arbitrary vector fields  we are
able to establish a polynomial structure in the second-order derivatives.
This structure is based on the some linear combinations of Olver 
hyper-Jacobians. We use as the main tools Fock space techniques and 
induction.  This structure can be used to analyze Lagrangian systems 
with groups of Noetherian symmetries. As an illustration we analyze 
the case of Lagrangian equations with Abelian gauge invariance.
\end{abstract}

\section{Introduction}

This paper is a continuation of \cite{G1}. The purpose is to completely
analyze the structure of locally variational second-order equations i.e.
second-order equations which can be obtained locally from some Lagrangian
function. The basic tool is, as in \cite {G1}, the generalization of the
Helmholtz-Sonin equations to field theory due to Anderson-Duchamp and Krupka
(ADK) \cite{AD}, \cite{Kr} and the mathematical framework for this formalism 
is the jet-bundle formalism.

In \cite {G1} we have succeeded to establish, for the particular case of a
scalar field, a polynomial structure in the second order derivatives for any
locally variational equations of second order. The main technical tricks were
some Fock space combinatorics together with the essential use of complete
induction. The result afforded a rather complete study of Lagrangian systems
with groups of Noetherian symmetries. In this paper we will be able to extend
the results from \cite {G1} to the case of arbitrary vector fields, and we
will prove that, again, a very specific polynomial structure in the
second-order derivatives appears. We will use the same combinatorial tricks
as in \cite{G1} and we fix a rather subtle point concerning some
identities satisfied by the hyper-Jacobians and the way to take
them into account.

In the mean time we became aware that result of this type have been also
obtained in \cite{A} by a rather different method. In this reference one can
find a conjecture on the structure of the most general case of locally
variational equations of arbitrary order. There are some differences
between our result and the results in \cite{A} which will be more convenient
to comment in the text of the paper after we will present the general strategy
of our approach. Nevertheless, the polynomial structure of a locally
variational equation of second-order is expressed in both approaches in
terms of the so-called hyper-Jacobians which 
have been introduced in the literature in \cite{BCO}, \cite{O}. We also 
mention \cite {AP1} where the interplay between locally variationallity and 
conservation laws for second order scalar equations is studied.

The paper is organized as follows. In Section 2 we remind the basic aspects of
the formalism following \cite{G1}. In Section 3 we derive our main result
concerning the polynomial structured expression of an arbitrary second-order
locally variational equation with an arbitrary number of fields. In Section 4 we
present a typical application as we have already anticipated. Namely, we will
study Lagrangian systems with groups of Noetherian symmetries. Our main concern
will be the case of infinite dimensional groups as Abelian gauge symmetry.
We will be able to show that in this cases a rather drastic simplification
occurs, i.e. the Euler-Lagrange expressions are polynomials of
first order in the second order derivatives and they can be obtained from
a first order (local) Lagrangian. In this way we are able to reobtain in a
completely different way a  result derived in \cite{G2}. We must
mention in this context the recent paper \cite{AP2} where the same type of
result concerning the Abelian invariance is obtained starting also from the
ADK equations but exploiting intensively the presence of some conservation
laws. 

\section{A Higher-Order Lagrangian Formalism}

2.1 The kinematical structure of classical field theory is based on a fibred
bundle structure
$
\pi: S \mapsto M
$
where
$S$
and
$M$
are differentiable manifolds of dimensions $
dim(M) = n,~dim(S) = N + n
$
and
$\pi$
is the canonical projection of the fibration. Usually $M$
is interpreted as the ``space-time" manifold and the fibres of $S$
as the field variables. Next, one considers the $r$-jet
bundle
$
J^{r}_{n}(S) \mapsto M \quad (r \in \N).
$
A $r$-order jet with source
$x \in U$,
and target
$y \in V$
is, by definition, an equivalence class of all the smooth maps
from
$
\zeta: M \rightarrow S
$
verifying
$\zeta(x) = y$
and having the same partial derivatives in $x$ up to order $r$ (in any
chart on $M$ and respectively on $S$).
We denote the equivalence class of
$\zeta$
by
$
j^{r}\zeta
$
and the factor set by
$
J^{r}_{x,y}.
$
Then the
$r$-order
jet bundle extension is, by definition
$
J^{r}_{n}(Y) \equiv \cup J^{r}_{x,y}.
$
The canonical projections
$
\pi^{s,t}: J^{s}_{n}(Y) \rightarrow J^{t}_{n}(Y)~~~(t \leq s \leq r)
$
are defined in a natural way. By convention
$
J^{0}_{n}(S) \equiv S
$
and
$
r \in \N \cup \{\infty\}.
$

One usually must take
$
r \in \N
$
but sufficiently large such that all formulas make sense. Let us consider a
local system of coordinates in the chart
$
U \subseteq S:~
(x^{\mu})~(\mu = 1,...,n).
$
Then on some chart
$
V \subseteq \pi^{-1}(U) \subset S
$
we take a local coordinate system adapted to the fibration structure: $
(x^{\mu},\psi^{A})~(\mu = 1,...,n,~A = 1,...,N) $
such that the canonical projection is
$
\pi(x^{\mu},\psi^{A}) = (x^{\mu}).
$

Then one can extend this system of coordinates to
$
J^{r}_{n}(S)
$
as follows: on the open set
$
V^{r} \equiv (\pi^{r,0})^{-1}(V)
$
we define the coordinates of
$
j^{r}_{x}\zeta
$
to be
$
(x^{\mu},\psi^{A},\psi^{A}_{\mu},...,\psi^{A}_{\mu_{1}.,,,,\mu_{r}})
$
where
$
\mu_{1} \leq \cdots \leq \mu_{s}~~~(s \leq r).
$
Explicitly

\be
\psi^{A}_{\mu_{1},...,\mu_{s}}(j^{r}_{x}\zeta) \equiv
\prod_{i=1}^{s} {\partial\over \partial x^{\mu_{i}}} \zeta(x) 
\quad (s=1,...,r).
\ee

If
$
\mu_{1},...,\mu_{s}
$
are arbitrary numbers belonging to the set
$
\{1,...,n\}
$
then by the expression
$
\{\mu_{1},...,\mu_{s}\}
$
we understand the result of the operation of increasing ordering. Then
the notation
$
\psi^{A}_{\mu_{1},...,\mu_{s}}
$
becomes meaningful for all set of numbers
$
\mu_{1},...,\mu_{s}.
$
If
$
I = \mu_{1},...,\mu_{s}
$
is an arbitrary set from
$
\{1,...,n\}^{\times s}
$
then we define

\be
\psi^{A}_{I} = \psi^{A}_{\mu_{1},...,\mu_{s}} \equiv
\psi^{A}_{\{\mu_{1},...,\mu_{s}\}}.
\ee

This notation makes sense whenever the cardinal of $I$ verifies:
$
|I| \leq r
$
where if
$
I = \emptyset
$
then we put
$
\psi^{A}_{\emptyset} = \psi^{A}.
$
With this convention the expression
$
\psi^{A}_{I}
$
is completely symmetric in the individual indices
$
\mu_{1},...,\mu_{s}
$
which make up the multi-index $I$.

2.2 Let us consider
$
s \leq r
$
and
$T$
a
$(n + 1)$-form
which can be written in the local coordinates introduced above as:

\be
T = {\cal T}_{A}~d\psi^{A} \wedge dx^{1} \wedge \cdots \wedge dx^{n}
\label{edif}
\ee
with
$
{\cal T}_{A}
$
some smooth functions of
$
(x^{\mu},\psi^{A}_{I})~~~(|I| \leq s).
$

Then
$T$
is a globally defined object. We call such a
$T$
a {\it differential equation of order s}.

2.3 To introduce some special type of differential equations we need some
very useful notations \cite{AD}. We define the differential operators:
\be
\partial^{I}_{A} \equiv {r_{1}!...r_{l}! \over |I|!}
{\partial \over \partial \psi^{A}_{I}}
\label{pdif}
\ee
where
$
r_{i}
$
is the number of times the index
$i$
appears in $I$.
The combinatorial factor in (\ref{pdif}) avoids possible overcounting in the
computations which will appear in the following. One has then:

\be
\partial^{\mu_{1},...,\mu_{l}}_{A} \psi^{B}_{\nu_{1},...,\nu_{m}} =
\cases{  { 1\over l!} \delta^{A}_{B}
perm(\delta^{\mu_{i}}_{\nu_{j}}), & for $l = m$ \cr
0,  & for $l \not= m$ \cr}
\ee
where

\be
perm\left( \delta^{\mu_{i}}_{\nu_{j}} \right) \equiv
\sum_{P \in {\cal P}_{l}} \delta^{\mu_{1}}_{\nu_{P(1)}}\cdots
\delta^{\mu_{l}}_{\nu_{P(l)}}
\ee
is a permanent. (In general we denote by
$
perm(A)
$
the permanent of the matrix
$A$).

Next, we define the total derivative operators:
\be
D_{\mu} = {\partial\over \partial x^{\mu}} + \sum_{l=0}^{r-1}
\psi^{A}_{\nu_{1},...,\nu_{l},\mu} \partial^{\nu_{1},...,\nu_{l}}_{A}
= {\partial\over \partial x^{\mu}} + \sum_{|I|\leq r-1} \psi^{A}_{I\mu}~
\partial^{I}_{A}
\label{tdif}
\ee
where we use the convention
$
IJ \equiv I \cup J.
$
One can check that
\be
D_{\mu}\psi^{A}_{I} = \psi^{A}_{I\mu}, \qquad |I| \leq r-1
\label{der}
\ee
and
\be
[D_{\mu}, D_{\nu}] = 0.
\label{com}
\ee

Finally we define the differential operators
\be
D_{I} \equiv \prod_{i \in I} D_{\mu_{i}}.
\label{tdifs}
\ee

Because of (\ref{com}) the order of the factors in the right hand side is
irrelevant.

2.4 A differential equation
$T$
is called {\it locally variational} (or of the {\it Euler-Lagrange type})
{\it iff} there exists a local real function
${\cal L}$
such that the functions
$
{\cal T}_{A}
$
from (\ref{edif}) are of the form:
\be
{\cal E}_{A}({\cal L}) \equiv \sum_{l=0}^{r} (-1)^{l}
D_{\mu_{1},...,\mu_{l}} (\partial^{\mu_{1},...,\mu_{l}}_{A} {\cal L})
\label{Eop}
\ee

One calls
${\cal L}$
a {\it local Lagrangian} and:

\be
L \equiv {\cal L}~dx^{1}\wedge\cdots \wedge dx^{n}
\label{Lform}
\ee
a {\it local Lagrange form}.

If the differential equation
$T$
is constructed as above then we denote it by
$
E(L).
$
A local Lagrangian is called a {\it total divergence} if it is of the form:
\be
{\cal L} = D_{\mu} V^{\mu}.
\ee

One can check that in this case we have:
\be
E(L) = 0.
\label{trEL}
\ee

2.5 We now state a central result of variational calculus which will be our
main tool in the following analysis (see \cite{AD}, \cite{Kr}):

\begin{thm}
Let
$T$
be a differential equation of order
$s$. Then
$T$
is locally variational {\it iff} the functions $
{\cal T}_{A}
$
from (\ref{edif}) verify the following equations:
\be
\partial^{\mu_{1},...,\mu_{l}}_{A} {\cal T}_{B} = \sum_{p=l}^{s} (-1)^{p}
C_{l}^{p} D_{\mu_{l+1},...,\mu_{p}} \partial^{\mu_{1},...,\mu_{p}}_{B}
{\cal T}_{A},~~~(l \leq s).
\label{ADK}
\ee
\label{AD}
The equations above  must be considered as differential equations of order
$
2s
$
with 
$
T_{A}
$
having a trivial dependence on the derivatives of order
$
s + 1,...,2s.
$
\end{thm}

These are the so-called {\it Anderson-Duchamp-Krupka equation}. We only
mention that if the equations above are fulfilled then a possible (local)
Lagrangian in given by the Tonti expression:

\be
{\cal L} = \int_{0}^{1} \psi^{A} {\cal T}_{A}\circ \chi_{\lambda} d\lambda
\label{Tonti}
\ee
where
$$
\chi_{\lambda} (x^{\mu},\psi^{A}_{I}) =
(x^{\mu},\lambda\psi^{A}_{I}).
$$

2.6  An {\it evolution} is any section
$
\Psi:M \rightarrow S
$
of the bundle
$
\pi: S \mapsto M.
$

Let us denote by
$
j^{s}\Psi:M \rightarrow J^{s}_{n}(S)
$
the natural lift of
$
\Psi
$
to a section of the
$s$-order
fibre bundle extension
$
\pi^{s,0}: J^{r}_{n}(S) \rightarrow M.
$
Then the action functional is defined by:
\be
{\cal A}_{L}(\Psi) \equiv \int (j^{s}\Psi)^{*} L.
\ee

Let us suppose that
$T$
is a differential equation and
$
\Psi: M \mapsto S
$
is a evolution. One says that
$
\Psi
$
is a {\it solution} of
$T$
if one has:
\be
{\cal T}_{A} \circ j^{s}\Psi = 0 \quad (A = 1,...,N).
\label{ELeq}
\ee

If
$T$
is locally variational
$
T = E(L)
$
one obtains the global form of the {\it Euler-Lagrange equations}. In local
coordinates one can arrange such that $
\Psi
$
has the form
$
x^{\mu} \mapsto (x^{\mu},\Psi(x));
$
then
$
j^{s} \Psi: M \rightarrow J^{s}_{n}(S)
$
is given by
$$
x^{\mu} \mapsto \left( x^{\mu},\prod_{i \in I} {\partial \over 
\partial x^{\mu_{i}}} \Psi (x) \right)
$$
and (\ref{ELeq}) take the well-known form.

By a {\it symmetry} of $T$ we understand a map
$
\phi \in Diff(S)
$
such that if
$
\Psi:M \rightarrow S
$
is a solution of
$T$,
then
$
\phi \circ \Psi
$
is a solution of
$T$
also. (In particular this definition assumes implicitly that
$
\phi \circ \Psi
$
is a evolution i.e. a section of
$
\pi: S \mapsto M
$.)

If one has
\be
(j^{s}\phi)^{*}~T = T,
\ee
where
$
j^{s}\phi \in Diff(J_{n}^{s}(S))
$
is the lift of
$
\phi
$,
then
$
\phi
$
is a symmetry. These type of symmetries are the well-known {\it Noetherian
symmetries}.

2.7 We particularize the ADK equations for case
$
s = 2
$
of second-order Euler-Lagrange equations. One obtains after some 
elementary computations the following set of equations:

\be
\partial_{A}^{\mu_{1}\mu_{2}} {\cal T}_{B} = \partial_{B}^{\mu_{1}\mu_{2}}
{\cal T}_{A}
\label{ADK1}
\ee

\be
\left(\partial_{B}^{\mu\rho_{1}}\partial_{C}^{\rho_{2}\rho_{3}} +
\partial_{B}^{\mu\rho_{2}}\partial_{C}^{\rho_{3}\rho_{1}} +
\partial_{B}^{\mu\rho_{3}}\partial_{C}^{\rho_{1}\rho_{2}}\right) {\cal T}_{A}
= 0
\label{ADK2}
\ee

\be
\partial_{A}^{\mu_{1}} {\cal T}_{B} +
\partial_{B}^{\mu_{1}} {\cal T}_{A} =
2\left(\partial_{\mu_{2}} + \psi^{C}_{\mu_{2}} \partial_{C} +
\psi^{C}_{\mu_{2}\nu_{1}} \partial_{C}^{\nu_{1}}\right)
\partial_{B}^{\mu_{1}\mu_{2}}{\cal T}_{A}
\label{ADK3}
\ee

\be
\begin{array}{c}
\partial_{A} {\cal T}_{B} - \partial_{B} {\cal T}_{A} =
-\left(\partial_{\mu_{1}} + \psi^{C}_{\mu_{1}} \partial_{C} +
\psi^{C}_{\mu_{1}\nu_{1}} \partial_{C}^{\nu_{1}}\right)
\partial_{B}^{\mu_{1}}{\cal T}_{A} +
\nonumber \\
\left(\partial_{\mu_{1}} + \psi^{C}_{\mu_{1}} \partial_{C} +
\psi^{C}_{\mu_{1}\nu_{1}} \partial_{C}^{\nu_{1}}\right)
\left(\partial_{\mu_{2}} + \psi^{D}_{\mu_{2}} \partial_{D} +
\psi^{D}_{\mu_{2}\nu_{2}} \partial_{D}^{\nu_{2}}\right)
\partial_{B}^{\mu_{1}\mu_{2}}{\cal T}_{A}.
\label{ADK4}
\end{array}
\ee

We will prove in the next Section that from (\ref{ADK1}) and (\ref{ADK2})
follows that
$
{\cal T}_{A}
$
has a certain polynomial structure in the second order derivatives
$
\psi^{A}_{\mu\nu}
$
i.e the dependence is through the second order hyper-Jacobians.

\begin{rem}

The result in \cite{A} seems to be that the emergence of the hyper-Jacobians
follows only from (\ref{ADK2}). Although this is true for a scalar field
(because (\ref{ADK1}) is trivial in this case), in the general this seems to
be wrong. We will start the induction process in the next Section with the case
$n = 2$
and the necessity of using both equations (\ref{ADK1}) and (\ref{ADK2}) to
obtain the hyper-Jacobians dependence will be clear. However it seems to be
true that a general polynomial structure can be obtained only from
(\ref{ADK2}).
\label{comparison}
\end{rem}

\section{The Main Theorem}

3.1 We turn now to the study of the equations (\ref{ADK1}) and (\ref{ADK2}).
Let us define the expressions:

\be
\psi^{A_{k+1},...,A_{n};\mu_{1},...,\mu_{k};\nu_{1},...,\nu_{k}} \equiv
{1\over (n-k)!}
\varepsilon^{\mu_{1},...,\mu_{n}} \varepsilon^{\nu_{1},...,\nu_{n}}
\prod_{i=k+1}^{n} \psi^{A_{i}}_{\mu_{i}\nu_{i}}~~(k = 0,...,n)
\label{minors}
\ee

We will use consistently Bourbaki conventions:
$
\sum_{\emptyset} \cdots = 0
$
and
$
\prod_{\emptyset} \cdots = 1.
$

These are some linear combinations of Olver hyper-Jacobians (see
\cite{BCO}, \cite{O}). We note the following symmetry properties:

\be
\psi^{A_{P(k+1)},...,A_{P(n)};\mu_{Q(1)},...,\mu_{Q(k)};
\nu_{R(1)},...,\nu_{R(k)}} = (-1)^{|Q|+|R|}
\psi^{A_{k+1},...,A_{n};\mu_{1},...,\mu_{k};\nu_{1},...,\nu_{k}}  
\quad (k = 0,...,n)
\label{antisym}
\ee
where
$P$ 
is a permutation of the numbers
$k+1,...,n$
and 
$Q, R$
are permutations of the numbers 
$1,...,k$
and

\be
\psi^{A_{k+1},...,A_{n};\nu_{1},...,\nu_{k};\mu_{1},...,\mu_{k}} =
\psi^{A_{k+1},...,A_{n};\mu_{1},...,\mu_{k};\nu_{1},...,\nu_{k}} 
\quad (k = 0,...,n).
\label{sym}
\ee

We also note that the following identities are true (see \cite{O}):

\be
\sum_{i=0}^{k} (-1)^{i} 
\psi^{A_{k+1},...,A_{n};\mu_{1},...,\mu_{k-1},\nu_{i};
\nu_{0},...,\hat{\nu_{i}},...,\nu_{k}} = 0.
\quad (k = 0,...,n).
\label{identities}
\ee

Let us first settle the following point: are the relations
(\ref{sym}) and (\ref{identities}) independent one of the other?
The answer is negative and follows from the following lemma:

\begin{lemma}
\label{symmetric}

If the tensor
$\psi$
verifies the identities (\ref{antisym}) and (\ref{identities})
then it also verifies:

\be
\begin{array}{c}
\sum_{i_{1}<\cdots <i_{p}} (-1)^{i_{1}+\cdots i_{p}} 
\psi^{A_{k+1},...,A_{n};\nu_{i_{1}},\cdots
\nu_{i_{p}},\mu_{1},...,\mu_{k-p};\mu_{k-p+1},\cdots
\mu_{k},\nu_{1},\cdots \hat{\nu_{i_{1}}}\cdot \hat{\nu_{i_{p}}}\cdots
\nu_{k}} =
\nonumber \\
=(-1)^{p(k-1)+p(p+1)/2} 
\psi^{A_{k+1},...,A_{n};\mu_{1},...,\mu_{k};\nu_{1},...,\nu_{k}}
\qquad p = 1,...,k.
\end{array}
\label{identities-p}
\ee
\end{lemma}

{\bf Proof:} 

By induction over 
$p$.
For 
$
p = 1
$
the relation (\ref{identities-p}) becomes
(\ref{identities}). After some combinatorial computations one
succeeds in going from 
$p$
to
$
p + 1.
$
$\nabla$

Now if we take into the formula above 
$
p = k
$
we obtain (\ref{sym}).

In the following we will need the expression for the derivatives
of the hyper-Jacobians. On easily finds out the following formula:
true:

\be
\begin{array}{c}
\partial^{\rho\sigma}_{B}
\psi^{A_{k+1},...,A_{n};\mu_{1},...,\mu_{k};\nu_{1},...,\nu_{k}}
= \nonumber \\
{1 \over 2(n-k)} \sum_{j=k+1}^{n} \delta_{B}^{A_{j}}
\left( \psi^{A_{k+1},...,\hat{A_{j}},...,A_{n};\rho,\mu_{1},...,\mu_{k};
\sigma,\nu_{1},...,\nu_{k}}\right) + (\rho \leftrightarrow \sigma)
\quad (k=0,...,n).
\label{derJ}
\end{array}
\ee

This formula suggests the use of the Fock space techniques. Let us emphasize
this point in detail. We will consider the functions
$
\psi^{A_{k},...,A_{};\mu_{1},...,\mu_{k};\nu_{1},...,\nu_{k}}
$
as the components of a tensor
$
\{\psi_{k}\} \in {\cal H} \equiv {\cal F}^{+}(\R^{N}) \otimes
{\cal F}^{-}(\R^{n}) \otimes {\cal F}^{-}(\R^{n})
$
where
$
\psi_{k}
$
belongs to the subspace of homogeneous tensors
$
{\cal H}_{n-k,k,k}
$
(where
$
{\cal H}_{p,q,t}
$
is the subspace of homogeneous tensors of degree
$
p, q
$
and $t$ respectively.) These tensors verify the following symmetry property:
\be
S\psi = \psi
\label{sympsi}
\ee
where
\be
S(\Lambda \otimes \phi \otimes \psi) = \Lambda \otimes \psi \otimes \phi,
~~~\forall \phi, \psi \in {\cal F}^{-}(\R^{n}),~~~\forall \Lambda \in
{\cal F}^{+}(\R^{N}).
\ee

We will denote by
$
e^{*}_{(A)}
$
and
$
e^{(A)}
$
the bosonic creation and respectively the annihilation operators acting in
$
{\cal F}^{+}(\R^{N});~
$
similarly we denote by
$
a^{*}_{(\alpha)}
$
and
$
a^{(\alpha)}
$
the fermionic creation and respectively the annihilation operators acting in
$
{\cal F}^{-}(\R^{n}).
$
We can define in a natural way bosonic and fermionic creation and
annihilation operators acting in
$
{\cal H}
$
by:
\be
f^{*}_{(A)} \equiv e^{*}_{(A)} \otimes {\bf 1} \otimes
{\bf 1},\qquad
b^{*}_{(\alpha)} \equiv {\bf 1} \otimes a^{*}_{(\alpha)} \otimes
{\bf 1}, \qquad
c^{*}_{(\alpha)} \equiv {\bf 1} \otimes {\bf 1} \otimes a^{*}_{(\alpha)}
\ee
and similarly for the annihilation operators.

With these notations one can rewrite (\ref{derJ}) in a more compact way,
namely:

\be
\partial^{\rho\sigma}_{A}~\psi_{k} = \alpha_{k}~f^{*}_{(A)}~
[b^{(\rho)} c^{(\sigma)} + b^{(\sigma)} c^{(\rho)}]~
\psi_{k+1};~~~\alpha_{k} \equiv {1 \over 2}
\times {1 \over (k+1) \sqrt{n-k}}~~~(k = 0,...,n)
\label{derJF}
\ee
where we use for simplicity the convention
$
{0 \over 0} = 0.
$

We need one more notation for our Fock space machinery, namely
$
<\cdot,\cdot>
$
which is the duality form between
$
{\cal H}
$
and
$
{\cal H}^{*}.
$

3.2 We prove now the main result.

\begin{thm}
The general solution of the equations (\ref{ADK1}) and (\ref{ADK2}) is of
the following form:

\be
{\cal T}_{A} = \sum_{k=0}^{n} {1\over (k!)^{2}}
t_{A,B_{k+1},...,B_{n};\mu_{1},...,\mu_{k};\nu_{1},...,\nu_{k}}
\psi^{B_{k+1},...,B_{n};\mu_{1},...,\mu_{k};\nu_{1},...,\nu_{k}}
\label{polyn}
\ee
where the functions
$
t_{...}
$
are independent of
$
\psi_{\mu\nu}^{B}
$:

\be
\partial^{\rho\sigma}_{B}
t_{A_{k},...,A_{n};\mu_{1},...,\mu_{k};\nu_{1},...,\nu_{k}} = 0
\quad (k = 0,...,n),
\ee
and have analogous properties as the hyper-Jacobians, namely the
(anti)symmetry property:

\be
t_{A_{P(k)},...,A_{P(n)};\mu_{Q(1)},...,\mu_{Q(k)};
\nu_{R(1)},...,\nu_{R(k)}} = (-1)^{|Q|+|R|}
t_{A_{k},...,A_{n};\mu_{1},...,\mu_{k};\nu_{1},...,\nu_{k}}  
\quad (k = 0,...,n)
\label{antisym-t}
\ee
(where
$P$ 
is a permutation of the numbers
$k,...,n$
and 
$Q, R$
are permutations of the numbers 
$1,...,k$) 
and also verify the identities:

\be
\sum_{i=0}^{k} (-1)^{i} 
t_{A_{k},...,A_{n};\mu_{1},...,\mu_{k-1},\nu_{i};
\nu_{0},...,\hat{\nu_{i}},...,\nu_{k}} = 0.
\quad (k = 1,...,n).
\label{identities-t}
\ee

The function coefficients
$
t_{...}
$
are uniquely determined by
$
{\cal T}_{A}
$
and the properties (\ref{antisym-t}) and (\ref{identities-t}) above.

\label{structure}
\end{thm}

{\bf Proof:}

It is similar to the one in \cite{G1}. However we have refined a little bit
our Fock space tools and filled some gaps in the previous proof. It is 
convenient to consider that
$
t_{A_{k},...,A_{n};\mu_{1},...,\mu_{k};\nu_{1},...,\nu_{k}}
$
are the components of a tensor
$
\{t^{k}\}
$
in the dual space
$
{\cal H}^{*};
$
explicitly:
$
t^{k} \in {\cal H}^{*}_{n-k+1,k,k}
$
(where
$
{\cal H}^{*}_{p,q,t}
$
is the subspace of homogeneous tensors of degree
$
p, q
$
and $t$ respectively.) Alternatively, it will be convenient to consider
$
t^{k}_{A}
$
as  a tensor from
$
{\cal H}^{*}_{n-k,k,k}
$
or
$
t_{A_{0},...,A_{k}}
$
as a tensor from
$
{\cal H}^{*}_{0,n-k,n-k}.
$

With this trick, formula (\ref{polyn}) can be written in
compact notations as:

\be
{\cal T}_{A} = \sum_{k=0}^{n} {1 \over (k!)^{2}} <t_{A}^{k},\psi_{k}>.
\label{compact}
\ee

We also note that from (\ref{antisym-t}) and (\ref{identities-t}) 
one can derive, using lemma \ref{symmetric} that the function
coefficients
$
t_{...}
$
verify also the symmetry property:

\be
t_{A_{k+1},...,A_{n};\nu_{1},...,\nu_{k};\mu_{1},...,\mu_{k}} =
t_{A_{k+1},...,A_{n};\mu_{1},...,\mu_{k};\nu_{1},...,\nu_{k}} 
\quad (k = 0,...,n).
\label{sym-t}
\ee

(i) We now prove the uniqueness statement. So we must show that if

\be
\sum_{k=0}^{n} {1 \over (k!)^{2}} <t_{A}^{k},\psi_{k}> = 0
\label{uniqueness}
\ee
then
$
t^{k} = 0.
$

To prove this, we apply to the equation (\ref{uniqueness}) the operator
$
\prod_{i=1}^{p} \partial^{\rho_{i}\sigma_{i}}_{A_{i}} \quad (p \leq n)
$
and then we will make
$
\psi^{A}_{\mu\nu} \rightarrow 0.
$
Using (\ref{derJF}) one easily discovers the following equations:

\be
\prod_{i=1}^{p}~\left[ b^{*(\rho_{i})} c^{*(\sigma_{i})} +
b^{*(\sigma_{i})} c^{*(\rho_{i})} \right]~t_{A_{0},...,A_{p}} = 0,
\qquad (p = 0,...,n).
\label{unicity}
\ee

To analyze this system we first define the operator:

\be
{\cal C} \equiv b_{(\rho)} c^{*(\rho)}
\ee
and prove by elementary computations that the condition
(\ref{identities-t}) can be rewritten as:

\be
{\cal C} t_{A_{k},...,A_{n}} = 0 \quad (k = 0,...,n).
\label{Ct}
\ee

We remark here that (\ref{sym-t}) can be written compactly as:

\be
S t_{A_{k},...,A_{n}} = t_{A_{k},...,A_{n}} \quad (k = 0,...n).
\label{symmm}
\ee

Because we have

\be
S {\cal C} = {\cal C}^{*} S
\ee
we get from (\ref{Ct}) and (\ref{symmm}) that one also has:

\be
{\cal C}^{*} t_{A_{k},...,A_{n}} = 0 \quad (k = 0,...,n).
\label{Ct-star}
\ee

Let us also define the number operators:

\be
N_{b} \equiv b^{*(\rho)} b_{(\rho)}; \quad 
N_{c} \equiv c^{*(\rho)} c_{(\rho)}.
\ee

Then one knows that:

\be
N_{b} \vert_{{\cal H}^{0}_{p,q}} = p {\bf 1}, \quad
N_{c} \vert_{{\cal H}^{0}_{p,q}} = q {\bf 1}.
\label{number}
\ee

We analyze the system (\ref{unicity}) using some simple lemmas.
The proofs are elementary and are omitted.

\begin{lemma}

The following formula is true:

\be
b_{(\rho)} c_{(\sigma)} \left[ b^{*(\rho)} c^{*(\sigma)} +
b^{*(\sigma)} c^{*(\rho)} \right] = \left[ (n+1) {\bf 1} - N_{b}
\right] (n {\bf 1} - N_{c}) - {\cal C}^{*} {\cal C}.
\ee
\label{inverse}
\end{lemma}

\begin{lemma}

The operator
${\cal C}$
commutes with all the operators of the form

$$
\left[ b^{*(\rho)} c^{*(\sigma)} + b^{*(\sigma)} c^{*(\rho)} \right].
$$ 

Explicitly:

\be
{\cal C} \left[ b^{*(\rho)} c^{*(\sigma)} +
b^{*(\sigma)} c^{*(\rho)} \right] = 
\left[ b^{*(\rho)} c^{*(\sigma)} +
b^{*(\sigma)} c^{*(\rho)} \right] {\cal C}.
\ee
\label{commute}
\end{lemma}

\begin{lemma}

If the tensor
$t$
verifies the identity
$
{\cal C} t = 0
$
the also the tensors
$$
\left[ b^{*(\rho)} c^{*(\sigma)} + b^{*(\sigma)} c^{*(\rho)}
\right] t
$$
verify this identity.
\label{iteration}
\end{lemma}

We now have:

\begin{prop}

Suppose the tensor
$
t \in {\cal H}^{0}_{n-p,n-p} \quad (p > 0)
$
verify the system:

\be
\prod_{i=1}^{p}~\left[ b^{*(\rho_{i})} c^{*(\sigma_{i})} +
b^{*(\sigma_{i})} c^{*(\rho_{i})} \right]~t = 0.
\ee

Then we have
$ 
t = 0.
$
\end{prop}

{\bf Proof:} We apply to the system above the operator
$
\prod_{i=1}^{p} b_{(\rho_{i})} c_{(\rho_{i})}
$
and make repeated use of the lemmas above.
$\nabla$

We will call the argument involved in the proof above {\it the
unicity argument}.

In conclusion the system (\ref{unicity}) has the solution
$
t_{A_{k},...,A_{n}} = 0 \quad (k = 0,...,n).
$

(ii) We start to prove the formula (\ref{polyn}) by induction over
$n$. For
$n = 1$
the derivation of (\ref{polyn}) is elementary.
We provide this analysis for
$
n = 2
$
also. In fact, the consideration of this case is the simplest way to guess
the right induction hypothesis and to see the appearance of the
hyper-Jacobians.
One discovers from (\ref{ADK2}) that
$
{\cal T}_{A}
$
is of the following form:
\be
{\cal T}_{A} = t_{A} + t^{\mu}_{A;B}~\psi^{B}_{\mu\mu} +
t_{A;B}~\psi^{B}_{12} + t_{A;BC}~(\psi^{B}_{11}\psi^{C}_{22} -
\psi^{B}_{12}\psi^{C}_{12})
\label{polyn2}
\ee
where the functions
$
t_{...}
$
do not depend on the second order derivatives. It is clear that the
hyper-Jacobians (\ref{minors}) are not present yet. But if we enforce 
(\ref{ADK1}) also, then we obtain that
$
t_{A;B}
$
is completely symmetric in $A$ and $B$; also that
$
t_{A;BC}
$
is completely symmetric in
$A, B$
and $C$. Simple relabelling transform now the expression (\ref{polyn2})
into the (\ref{polyn}) for
$
n = 2.
$
We also note that the identities (\ref{identities-t}) are
satisfied in this case, so the step 
$
n = 2
$
of the induction procedure is completed.

(iii) We suppose that we have the assertion of the theorem for a given $n$
and we prove it for
$
n + 1.
$
In this case the indices
$
\mu,\nu, ...
$
takes values (for notational convenience)
$
\mu,\nu, ...= 0,...,n
$
and
$
i,j,...= 1,...,n.
$

If we consider in (\ref{ADK1}) and (\ref{ADK2}) that
$
\mu,\rho_{1},\rho_{2},\rho_{3} = 1,...,n
$
then we can apply the induction hypothesis and we get.
\be
{\cal T}_{A} = \sum_{k=0}^{n} {1\over (k!)^{2}}
\tilde{\cal T}_{AB_{k+1},...,B_{n};i_{1},...,i_{k};j_{1},...,j_{k}}
\tilde{\psi}^{B_{k+1},...,B_{n};i_{1},...,i_{k};j_{1},...,j_{k}}
\label{polyn'}
\ee

Here
$
\tilde{\cal T}_{...}
$
has properties of the type
(\ref{antisym-t})-({\ref{identities-t}) and can depend on
$
x, \psi^{A}, \psi^{A}_{\mu}
$
{\it and}
$
\psi^{A}_{0\mu}.
$
The expressions
$
\tilde{\psi}^{...}
$
are constructed from
$
\psi^{A}_{ij}
$
according to the prescription (\ref{minors}).

(iv) We still have at our disposal the relations (\ref{ADK1}) and
(\ref{ADK2}) where at least one index takes the value $0$. The computations
are rather easy to do using instead of (\ref{polyn'}) the compact tensor
notation (see (\ref{compact})) and the unicity argument. We obtain rather 
easily from (\ref{ADK1}):

\be
\partial^{0\mu}_{B}~\tilde{\cal T}_{A_{k},...,A_{n}} =
(A_{k} \leftrightarrow B) \qquad (k = 0,...,n)
\label{eqI}
\ee

and from (\ref{ADK2}):

\be
\partial^{00}_{B} \partial^{00}_{C}
\tilde{\cal T}_{A_{k},...,A_{n}} = 0,
\qquad (k = 0,...,n)
\label{eq1}
\ee

\be
\partial^{00}_{B}\partial^{0l}_{C}
\tilde{\cal T}_{A_{k},...,A_{n}} = 0,
\qquad (k = 0,...,n)
\label{eq2}
\ee

\be
\partial^{0l}_{B}\partial^{0m}_{C}
\tilde{\cal T}_{A_{0},...,A_{n}} = 0
\label{eq3}
\ee

\be
{1\over 2}k \left[ b^{*(l)} c^{*(m)} + b^{*(m)} c^{*(l)} \right]
\quad \partial^{00}_{C}~\tilde{\cal T}_{B,A_{k},...,A_{n}} +
2 \partial^{0l}_{B} \partial^{0m}_{C}~
\tilde{\cal T}_{A_{k},...,A_{n}} = 0, \qquad (k = 1,...,n)
\label{eq4}
\ee

\be
\sum_{(l,m,r)} \left[ b^{*(l)} c^{*(m)} + b^{*(m)} c^{*(l)} \right]
\quad \partial^{0r}_{B} \tilde{\cal T}_{A_{k},...,A_{n}} = 0 \qquad
(\forall k = 1,...,n-1).
\label{eq5}
\ee

Here by
$
\sum_{(l,m,r)}
$
we understand the sum over all cyclic permutations of the indices
$
l,m,r.
$

The expressions
$
\tilde{\cal T}_{A_{k},...,A_{n}}
$
are obviously considered as tensors from
$
{\cal H}^{*}_{0,k,k}
$
verifying the restriction:

\be
S  \tilde{\cal T}_{A_{k},...,A_{n}} = \tilde{\cal T}_{A_{k},...,A_{n}}
\quad (k = 0,...,n)
\label{symrestr}
\ee

and

\be
\tilde{\cal C} \tilde{\cal T}_{A_{k},...,A_{n}} = 0
\quad (k =0,...,n)
\label{id1}
\ee
where we have defined:

\be
\tilde{\cal C} \equiv b_{(i)} c^{*(i)}.
\ee

One can easily show that from this identity we also have:

\be
{\cal C}^{*} \tilde{\cal T}_{A_{k},...,A_{n}} = 0
\quad (k =0,...,n).
\label{id2}
\ee

As in \cite{G1}, these equations  can be solved i.e. one can describe the
most general solution.

From (\ref{eq1}) we have:
\be
\tilde{\cal T}_{A_{k},...,A_{n}} = {\cal T}^{(0)}_{A_{k},...,A_{n}} +
\psi^{B}_{00} {\cal T}^{(1)}_{B;A_{k},...,A_{n}} \qquad (k = 0,...,n)
\label{sol}
\ee
where the functions
$
{\cal T}^{(0)}_{A_{k},...,A_{n}}
$
and
$
{\cal T}^{(1)}_{B;A_{k},...,A_{n}}
$
do not depend on the second order derivatives, are completely symmetric in
the indices
$
A_{k},...,A_{n}
$
and verify relations of the type (\ref{symrestr}), (\ref{id1})
and (\ref{id2}). However, if we use the relation (\ref{eqI}) for
$
\mu = 0
$
we get that in fact the tensor
$
{\cal T}^{(1)}_{B;A_{k},...,A_{n}}
$
is completely symmetric in {\it all} indices
$
B;A_{k},...,A_{n}.
$
So, we can rewrite (\ref{sol}) in a more precise way:

\be
\tilde{\cal T}_{A_{k},...,A_{n}} = {\cal T}^{(0)}_{A_{k},...,A_{n}} +
\psi^{B}_{00} {\cal T}^{(1)}_{B,A_{k},...,A_{n}} \qquad (k = 0,...,n)
\label{solI}
\ee
where the tensors
$
{\cal T}^{(1)}_{B,A_{k},...,A_{n}}
$
is completely symmetric in all indices; in other words
$
{\cal T}^{(0)}_{k} \in {\cal H}^{*}_{n-k+1,k,k}
$
and
$
{\cal T}^{(1)}_{k} \in {\cal H}^{*}_{n-k+2,k,k}.
$

We also have independence of {\it all} the second order derivatives for these
tensors and properties of the type (\ref{symrestr}), (\ref{id1})
and (\ref{id2}).

From (\ref{eq2}) and (\ref{eq3}) we also get:
\be
\partial^{0l}_{B} {\cal T}^{(1)}_{A_{k-1},...,A_{n}} = 0, \qquad
(k = 0,...,n)
\label{restr1}
\ee
and

\be
\partial^{0l}\partial^{0m} {\cal T}^{(0)}_{A_{0},..,A_{n}} = 0.
\label{restr3}
\ee

Equation (\ref{eqI}) for
$
\mu = l
$
gives:

\be
\partial^{0l}_{B}\quad \tilde{\cal T}^{(0)}_{A_{k},...,A_{n}} =
(A_{k} \leftrightarrow B) \qquad (k = 0,...,n)
\label{eqI'}
\ee

and finally (\ref{eq4}) and (\ref{eq5}) become:

\be
{1\over 2}k \left[ b^{*(l)} c^{*(m)} + b^{*(m)} c^{*(l)} \right]
{\cal T}^{(1)}_{B,C,A_{k},...,A_{n}} +
2 \partial^{0l}_{B} \partial^{0m}_{C}
{\cal T}^{(0)}_{A_{k},...,A_{n}} = 0, \qquad (k = 1,...,n)
\label{eq4'}
\ee

\be
\sum_{(l,m,r)} \left[ b^{*(l)} c^{*(m)} + b^{*(m)} c^{*(l)} \right]
\quad \partial^{0r} {\cal T}^{(0)}_{A_{k},...,A_{n}} = 0 \qquad
(\forall k = 1,...,n).
\label{eq5'}
\ee

(v) We proceed further by applying the operator
$
\partial^{0r}
$
to (\ref{eq4'}); taking into account (\ref{restr1}) we obtain:

\be
\partial^{0r}_{B} \partial^{0l}_{C} \partial^{0m}_{D}
{\cal T}^{(0)}_{A_{k},...,A_{n}} = 0 \qquad (k = 1,...,n)
\label{eq4''}
\ee

From (\ref{restr3}) and (\ref{eq4''}) one obtains a polynomial structure in
$
\psi^{B}_{0l}
$
for
$
{\cal T}^{(0)}_{...}.
$

\be
{\cal T}^{(0)}_{A_{0},...,A_{n}} =
{\cal T}^{(00)}_{A_{0},...,A_{n}} + {\cal T}^{l}_{B;A_{0},...,A_{n}}
\psi^{B}_{0l}
\label{sol1}
\ee
and

\be
{\cal T}^{(0)}_{A_{k},...,A_{n}} = {\cal T}^{(00)}_{A_{k},...,A_{n}} +
{\cal T}^{l}_{B;A_{k},...,A_{n}} \psi^{B}_{0l} +
{1\over 2} {\cal T}^{lm}_{B,C;A_{k},...,A_{n}}
\psi^{B}_{0l}~\psi^{C}_{0m}, \qquad (k = 1,...,n)
\label{sol2}
\ee
where the coefficients are tensors independent of the second order
derivatives and completely symmetric in the indices
$
A_{k},...,A_{n}.
$
Moreover, we can assume that:

\be
{\cal T}^{lm}_{B,C;A_{k},...,A_{n}} =\quad  
{\cal T}^{ml}_{C,B;A_{k},...,A_{n}}.
\ee

One can obtain more information if one plugs these equations into
(\ref{eqI'}): as one can expect by now we get full symmetry in all indices
$
B, A_{k},...,A_{n}
$
and respectively
$
B,C, A_{k},...,A_{n}.
$

So, finally we have the following polynomial structure:

\be
{\cal T}^{(0)}_{A_{0},...,A_{n}} =
{\cal T}^{(00)}_{A_{0},...,A_{n}} + {\cal T}^{l}_{B,A_{0},...,A_{n}}
\psi^{B}_{0l}
\label{sol1'}
\ee
and

\be
{\cal T}^{(0)}_{A_{k},...,A_{n}} = {\cal T}^{(00)}_{A_{k},...,A_{n}} +
{\cal T}^{l}_{B,A_{k},...,A_{n}} \psi^{B}_{0l} +
{1\over 2} {\cal T}^{lm}_{B,C,A_{k},...,A_{n}}
\psi^{B}_{0l}~\psi^{C}_{0m}, \qquad (k = 1,...,n)
\label{sol2'}
\ee
with the tensors
$
{\cal T}^{(00)}_{...}, \quad {\cal T}^{l}_{...}
$
and
$
{\cal T}^{lm}_{...}
$
completely symmetric in the lower indices or, in different phrasing
$
{\cal T}^{(00)}_{k} \in {\cal H}^{*}_{n-k+1,k,k},
~{\cal T}^{l}_{k} \in {\cal H}^{*}_{n-k+2,k,k}
$
and
$
{\cal T}^{lm}_{K} \in {\cal H}^{*}_{n-k+3,k,k}.
$

These expressions also verify the symmetry property

\be
S {\cal T}^{...} = {\cal T}^{..},
\label{symmpropr}
\ee
the independence property of {\it all} of the second order derivatives
and the identities:

\be
\tilde{\cal C} {\cal T}^{...} = 0 \qquad
\tilde{\cal C}^{*} {\cal T}^{...} = 0.
\label{tilda-id}
\ee 

We can also suppose that:
\be
{\cal T}^{lm}_{A_{k-2},...,A_{n}} =\quad {\cal T}^{ml}_{A_{k-2},...,A_{n}}.
\ee

It remains to use the equations (\ref{eq4'}) and (\ref{eq5'}).
If we insert (\ref{sol2'}) into (\ref{eq4'}) we get:

\be
{\cal T}^{lm}_{A_{k-2},...,A_{n}} = - {1\over 2}~k
\left[ b^{*(l)} c^{*(m)} + b^{*(m)} c^{*(l)} \right]
{\cal T}^{(1)}_{A_{k-2},...,A_{n}} \qquad (k = 0,...,n-1).
\label{sol3}
\ee

One must check that this identity is compatible with (\ref{tilda-id})
by applying the operators
$
\tilde{\cal C}
$
and
$
\tilde{\cal C}^{*}
$
to this relation.

Finally, inserting (\ref{sol1'}) and (\ref{sol2'}) into (\ref{eq5'}) we get:
\be
\sum_{(l,m,r)} \left[ b^{*(l)} c^{*(m)} + b^{*(m)} c^{*(l)} \right]
{\cal T}^{r}_{A_{k-1},...,A_{n}} = 0\qquad (k = 1,...,n-1)
\label{eq6}
\ee
and

\be
\sum_{(l,m,r)} \left[ b^{*(l)} c^{*(m)} + b^{*(m)} c^{*(l)} \right]
{\cal T}^{rs}_{A_{k-2},...,A_{n}} = 0 \qquad (k = 1,...,n-1).
\label{eq7}
\ee

In conclusion the solution of (\ref{eqI})-(\ref{eq5}) is given by (\ref{sol})
where
$
{\cal T}^{(0)}
$
is given by (\ref{sol1'}) and (\ref{sol2'}); the independence of the
second-order derivatives have to be taken into account, and we are left
with the equations (\ref{sol3}), (\ref{eq6}) and (\ref{eq7}).

One shows immediately that (\ref{sol3}) identically verifies
(\ref{eq7}) so in fact we are left to solve only
(\ref{eq6}) together with the restrictions (\ref{symmpropr}),
(\ref{id1}) and (\ref{id2}). We have the following result:

\begin{lemma}

Let
$
T^{r} \in {\cal H}^{*}_{o,p,q} \qquad (p, q \leq n)
$
verifying the restrictions

\be
{\cal C} T^{r} = 0, \quad {\cal C}^{*} T^{r} = 0 
\label{id1 and id2}
\ee
and the system:

\be
\sum_{(l,m,r)}\left[ b^{*(l)} c^{*(m)} + b^{*(m)} c^{*(l)}
\right] \quad T^{r} = 0
\label{permutation}
\ee
for all 
$
l, r, m = 1,...,n.
$

Then one can write {\it uniquely}
$
T^{r}
$
in of the following form:

\be
T^{r} = b^{*(r)} T_{1} + c^{*(r)} T_{2}
\label{T}
\ee
with
$
T_{1} \in {\cal H}^{*}_{0,p-1,q}
$
and
$
T_{2} \in {\cal H}^{*}_{0,p,q-1}.
$

If
$
p = q
$
then
$
T_{2} = S T_{1}
$
\label{perm}
and

\be
\tilde{\cal C} T_{1} = 0, \qquad 
\tilde{\cal C} T_{2} = T_{1}.
\label{T-id}
\ee

\end{lemma}

{\bf Proof:}
We apply to the equation (\ref{permutation}) the operator
$
b_{(l)} c_{(m)}
$
and we find out (after summation over $l$ and $m$ and taking
into account (\ref{id1 and id2}):

\be
(n-p+2)(n-q+1) T^{r} = (n-p+1) b^{(r)} b_{(l)} T^{l} + (n-p+1)
c^{(r)} c_{(l)} T^{l}.
\ee

So we have the formula from the statement with:
$
T_{1} = (n-p+2)^{-1} b_{(l)} T^{l}
$
and
$
T_{2} = (n-p+2)^{-1} (n-q+1)^{-1} (n-p+1) c_{(l)} T^{l}.
$

It is obvious that for
$
p = q
$
we have
$
T_{2} = S T_{1}
$
and also (\ref{T-id}).

It remains to check that the equation (\ref{permutation}) is
indeed identically verified by (\ref{T}).
$\nabla$

Using again index notations, it follows that the most general solution of
(\ref{eq6}) is of the form:

\be
\begin{array}{c}
{\cal T}^{l}_{A_{k-1},...,A_{n};i_{1},...,i_{k};j_{1},...,j_{k}}
= 
\nonumber \\
{1 \over \sqrt{k}} \sum_{p=1}^{k} (-1)^{p-1}
(\delta_{i_{p}}^{l}
{\cal T}_{A_{k-1},...,A_{n};i_{1},...,\hat{i_{p}},...i_{k};j_{1},...,j_{k}} +
\delta_{j_{p}}^{l}
{\cal T}_{A_{k-1},...,A_{n};j_{1},...,\hat{j_{p}},...,j_{k};i_{1},...,i_{k}})
\nonumber \\(k = 1,...,n)
\label{sol6k}
\end{array}
\ee
and

\be
{\cal T}^{l}_{A_{-1},...,A_{n}} = 0.
\label{sol60}
\ee

The expression of
$
{\cal T}^{lm}_{...}
$
in terms of
$
{\cal T}^{(1)}_{...}
$
is

\be
\begin{array}{c}
{\cal T}^{lm}_{A_{k-2},...,A_{n};i_{1},...,i_{k};j_{1},...,j_{k}} =
\nonumber \ \sum_{p=1}^{k} (-1)^{p+q}
\left( \delta_{i_{p}}^{l} \delta_{j_{q}}^{m} +
      \delta_{i_{p}}^{m} \delta_{j_{q}}^{l}\right) {\cal
T}^{(1)}_{A_{k-2},...,A_{n};i_{1},...,\hat{i_{p}},...i_{k};j_{1},...,j_{k}}
\quad (k = 1,...,n).
\label{sol7}
\end{array}
\ee
where
$
{\cal T}_{A_{k-1},...,A_{n}} \in {\cal H}^{*}_{0,k-1,k}
$
or more precisely, they are the components of a tensor
$
{\cal T}_{k} \in {\cal H}^{*}_{n-k+1,k-1,k}.
$

This tensors are not completely arbitrary. Because of (\ref{T-id})
they must satisfy the following relations:

\be
\tilde{\cal C} {\cal T}_{A_{k-1},...,A_{n}} = 0, \quad
\tilde{\cal C} S {\cal T}_{A_{k-1},...,A_{n}} = 
{\cal T}_{A_{k-1},...,A_{n}} = 0 \quad (k = 0,...,n).
\label{idenT}
\ee

The structure of
$
{\cal T}_{i_{1},...,i_{k};j_{1},...,j_{k}}
$
is completely elucidated.

(vi) It remains to introduce these expressions for
$
{\cal T}_{A_{k+1},...,A_{n}}
$
in (\ref{polyn'}) and regroup the terms. Like in \cite{G1} one obtains the
desired formula (\ref{polyn}) for
$
n+1
$
with the tensors
$
t_{...}
$
expressed in terms of the tensors
$
\tilde{\cal T}^{...}_{...}
$
defined in the proof above. Finally one must check that these
new tensors
$
t_{...}
$
also verify the induction hypothesis i.e. the identities (\ref{Ct})
and (\ref{symmm}). This is done after some computations using (\ref{idenT})
and the induction is finished.
\qed

\begin{rem}

In the usual cases the number of field components $N$ is dependent on the
dimension of space-time $n$. For instance,
$
N = n
$
for a vector field,
$
N = n \times dim(G)
$
for a gauge theory of the Lie group $G$,
$
N = n(n+1)/2
$
for the gravitational field, etc. One may wonder if the theorem above applies
in this case also. That this is indeed so can be seen as follows. In the
theorem above, $N$ was always fixed, but {\it arbitrary}. So, we have in fact
proved, by induction, a whole host of theorems
$
T_{N,n}~(n = 1,2,...)
$
for
$
N = 1,2....
$
The cases of physical interest above are some sort of ``diagonal" theorems
$
T_{f_{n},n}.
$
\end{rem}

3.3 We can insert the solution (\ref{polyn}) of (\ref{ADK1}) and (\ref{ADK2})
into (\ref{ADK3}) and (\ref{ADK4}) to obtain further restrictions on the
functions
$
t_{...}.
$

It is convenient to define:
\be
{\delta\over\delta x^{\mu}} \equiv {\partial\over\partial x^{\mu}} +
\psi^{A}_{\mu} \partial_{A}.
\ee

The idea is to use the compact notation (\ref{compact}) and to introduce it in
the remaining ADK equations. We will use the same procedure as the one
used to prove the uniqueness statement from the main theorem, i.e. we will
apply to these equations the operators
$
\prod_{i=1}^{p} \partial^{\mu_{i}\nu_{i}}_{C_{i}}~(p \leq n)
$
and afterwards we will take
$
\psi_{\mu\nu}^{A} = 0.
$

We give only the final result. From (\ref{ADK3}) one gets:

\begin{equation}
\begin{array}{c} 
\label{ADK5} 
\prod_{i=1}^{p}
\left[ b^{*(\rho_{i})} c^{*(\sigma_{i})} +
b^{*(\sigma_{i})} c^{*(\rho_{i})} \right]
( \partial^{\mu}_{A} t_{B,C_{1},...,C_{p}} +
\partial^{\mu}_{B} t_{A,C_{1},...,C_{p}} -
\nonumber \\
(n-p) \left[ b^{*(\mu)} c^{*(\nu)} + b^{*(\nu)} c^{*(\mu)} \right]
{\delta \over \delta x^{\nu}} t_{A,B,C_{1},...,C_{p}} ) =
\sum_{j=1}^{p} \prod_{i\not= j}
\nonumber \\
\left[ b^{*(\rho_{i})} c^{*(\sigma_{i})} +
b^{*(\sigma_{i})} c^{*(\rho_{i})} \right]
\left( \left[ b^{*(\mu)} c^{*(\sigma_{j})} +
b^{*(\sigma_{j})} c^{*(\mu)} \right]
\partial^{\nu_{j}}_{C_{j}} t_{A,B,C_{1},...,\hat{C_{j}},...,C_{p}}
+ (\rho_{j} \leftrightarrow \sigma_{j}) \right)
\end{array}
\end{equation}

for
$
p = 0,...,n
$
and from (\ref{ADK4}) only the antisymmetric contribution in $A$ and $B$
gives something new, namely:

\begin{equation}
\begin{array}{c}
\label{ADK6} 
\prod_{i=1}^{p}
\left[ b^{*(\rho_{i})} c^{*(\sigma_{i})} +
b^{*(\sigma_{i})} c^{*(\rho_{i})} \right]
\left[2 \partial_{B} t_{A,C_{1},...,C_{p}}
- {\delta \over \delta x^{\nu}} \partial^{\nu}_{B} t_{A,C_{1},...,C_{p}}
- (A \leftrightarrow B) \right] =
\nonumber \\
{1 \over n-p+1} \sum_{j=1}^{p} \prod_{i\not=j}
\left[ b^{*(\rho_{i})} c^{*(\sigma_{i})} +
b^{*(\sigma_{i})} c^{*(\rho_{i})} \right]
\left[\left( \partial^{\rho_{j}}_{C_{j}} \partial^{\sigma_{j}}_{B} +
\partial^{\sigma_{j}}_{C_{j}} \partial^{\rho_{j}}_{B} \right)
t_{A,C_{1},...,\hat{C_{j}},...,C_{p}} -
(A \leftrightarrow B) \right] 
\nonumber \\
\quad (p = 0,...,n).
\end{array}
\end{equation}

\begin{rem}

For the case
$
N = 1
$
(\ref{ADK6}) is trivial and one can show that (\ref{ADK5}) can be 
written in the equivalent form 

\be
2 \partial^{\rho} t_{k} - k \left[ b^{*(\rho)} c^{*(\sigma)} +
b^{*(\sigma)} c^{*(\rho)} \right] {\delta \over \delta x^{\sigma}} 
t_{k-1} = \left[ b_{(\sigma)} b^{*(\rho)} + c_{(\sigma)} c^{*(\rho)}
\right] \partial^{\rho} t_{k} \quad (k = 0,...,n).
\label{ADK7}
\ee

Indeed, one obtains (\ref{ADK7}) from (\ref{ADK5}) by
applying the operator
$
\prod_{i=1}^{p} b_{(\rho_{i})} c_{(\sigma_{i})} 
$
and using lemmas 1-3. In the opposite direction, one applies
$
\prod_{i=1}^{p}
\left[ b^{*(\rho_{i})} c^{*(\sigma_{i})} +
b^{*(\sigma_{i})} c^{*(\rho_{i})} \right]
$
to (\ref{ADK7}) and after some manipulations one gets (\ref{ADK5}).

The identity (\ref{ADK7}) coincides with the result from \cite{G1}.
We also note that in the case of a Poincar\'e invariant scalar
field one can prove that the restrictions (\ref{identities-t})
are identically satisfied.
\end{rem}

3.4 We now study a particular case which will be very important for the
applications from the next section, namely the case when:
\be
t_{A_{0},...,A_{p}} = 0 \qquad (p > 1).
\label{first-order}
\ee

Among other things, this is the only case when we stand a chance to derive
the Euler-Lagrange expressions from a first-order Lagrangian, because if we
have (\ref{Eop}) with the Lagrangian of the first-order, then necessarily the
expressions
$
{\cal T}_{A}
$
are at most linear in the second order derivatives.

In the case the relation (\ref{first-order}) is true it is more convenient to
introduce the following notations:

\be
\tilde{t}_{A} \equiv {1 \over (n!)^{2}}
\varepsilon^{\mu_{1},...,\mu_{n}} \varepsilon^{\nu_{1},...,\nu_{n}}
t_{A;\mu_{1},...,\mu_{n};\nu_{1},...,\nu_{n}}
\ee

and

\be
\tilde{t}^{\mu_{1}\nu_{1}}_{AB} \equiv {1 \over ((n-1)!)^{2}}
\varepsilon^{\mu_{1},...,\mu_{n}} \varepsilon^{\nu_{1},...,\nu_{n}}
t_{A,B;\mu_{2},...,\mu_{n};\nu_{2},...,\nu_{n}}.
\ee

We note the symmetry property:

\be
\tilde{t}^{\mu\nu}_{AB} = \tilde{t}^{\nu\mu}_{AB} = \tilde{t}^{\mu\nu}_{BA}
\ee
and the fact that one can rewrite the Euler-Lagrange expressions as follows:

\be
{\cal T}_{A} = \tilde{t}^{\mu\nu}_{AB}~\psi^{B}_{\mu\nu} + \tilde{t}_{A}.
\label{Eop-second}
\ee

One can easily write the equations (\ref{ADK5}) and (\ref{ADK6}) in this
case. The result is:

\be
\partial^{\mu}_{A} \tilde{t}_{B} + \partial^{\mu}_{B} \tilde{t}_{A} =
2~{\delta \over \delta x^{\nu}} \tilde{t}^{\mu\nu}_{AB},
\label{1}
\ee

\be
\partial^{\rho}_{C} \tilde{t}^{\mu\nu}_{AB} +
\partial^{\rho}_{B} \tilde{t}^{\mu\nu}_{AC} =
\partial^{\mu}_{A} \tilde{t}^{\rho\nu}_{BC} +
\partial^{\nu}_{A} \tilde{t}^{\rho\mu}_{BC},
\label{2}
\ee

\be
2 \partial_{B} \tilde{t}_{A} - {\delta \over \delta x^{\nu}}
\partial^{\nu}_{B} \tilde{t}_{A} = A \leftrightarrow B,
\label{3}
\ee

\be
4 \partial_{B} \tilde{t}^{\mu\nu}_{AC} - 2 {\delta \over \delta x^{\rho}}
\partial^{\rho}_{B} \tilde{t}^{\mu\nu}_{AC} -
(\partial^{\mu}_{B}  \partial^{\nu}_{C} +
\partial^{\nu}_{B}  \partial^{\mu}_{C}) \tilde{t}_{A} =
A \leftrightarrow B,
\label{4}
\ee

\be
(\partial^{\mu}_{B}  \partial^{\nu}_{C} +
\partial^{\nu}_{B}  \partial^{\mu}_{C})  \tilde{t}^{\rho\sigma}_{AD} =
(\partial^{\rho}_{A}  \partial^{\sigma}_{D} +
\partial^{\sigma}_{A}  \partial^{\rho}_{D}) \tilde{t}^{\mu\nu}_{BC}.
\label{5}
\ee

From (\ref{Eop-second}) and the expression of the Tonti Lagrangian
(\ref{Tonti}) we obtain in this particular case:

\be
{\cal L} = {\cal L}_{0} + {\cal L}_{1}
\label{Tonti-second}
\ee
where
\be
{\cal L}_{0} \equiv \int_{0}^{1} \psi^{A} \tilde{t}_{A} \circ \chi_{\lambda}
d\lambda,
\ee
and
\be
{\cal L}_{1} \equiv \psi^{A}_{\mu\nu} {\cal L}^{\mu\nu}_{A}
\ee
with
\be
{\cal L}^{\mu\nu}_{A} = \int_{0}^{1} \lambda \psi^{B} \tilde{t}^{\mu\nu}_{AB}
\circ \chi_{\lambda} d\lambda
\ee
and
\be
\chi_{\lambda} (x^{\mu}, \psi^{A}, \psi^{A}_{\nu}) =
(x^{\mu},\lambda \psi^{A}, \lambda\psi^{A}_{\nu}).
\ee

3.5 We will need  the following result (see \cite{AD}):

\begin{thm}

The Lagrangian (\ref{Tonti-second}) is equivalent to a first-order
Lagrangian.
\label{reducibility}
\end{thm}

We name this result the order reduction theorem. We only mention that
the proof uses only the relation (\ref{2}) and (\ref{5}); more precisely,
two consequences of these relations expressed in relations verified by
the functions
$
{\cal L}^{\mu\nu}_{A}.
$

\begin{cor}

If the relations (\ref{2}) and (\ref{5}) are true, then there exists a
(local) function
${\cal L}$
independent of the second-order derivatives

\be
\partial^{\mu\nu}_{A} {\cal L} = 0
\ee

such that

\be
\tilde{t}^{\rho\sigma}_{AB} = -{1 \over 2} \left(
\partial^{\rho}_{A} \partial^{\sigma}_{B} +
\partial^{\sigma}_{A} \partial^{\rho}_{B} \right) {\cal L}
\ee
and

\be
\tilde{t}_{A} = \partial_{A} {\cal L} - {\delta \over \delta x^{\mu}}
\partial^{\mu}_{A} {\cal L}.
\ee

The function
${\cal L}$
is determined by the first relation above up to an expression of the
following form:

\be
{\cal L} = \sum_{k=0}^{n} {1\over k!}
C^{\mu_{1},...,\mu_{k}}_{A_{1},...,A_{k}}
\prod_{I=0}^{k} \psi^{A_{i}}_{\mu_{i}}
\ee
where the functions
$
C^{\mu_{1},...,\mu_{k}}_{A_{1},...,A_{k}}
$
are independent of the second order-derivatives and are completely
antisymmetric in the indices
$
\mu_{1},...,\mu_{k}
$
and in the indices
$
A_{1},...,A_{k}.
$
\end{cor}

The first part of the corollary is a reformulation of the reduction of the
order theorem stated above and the second assertion is elementary to prove
and it is the first step in deriving the most general trivial first-order
Lagrangian.

\section{Lagrangian Systems with Groups of Noetherian
Symmetries}

4.1 We consider only the case of an Abelian gauge theory without matter
fields. In the general framework of Section 2, we take
$
S=M \times M
$
where $M$ is the
$n$-dimensional
Minkowski space with global coordinates
$
(x^{\mu},A_{\lambda})
$
(here
$
\mu,\nu=1,...,n)
$
and
$
X = M
$
with global coordinates
$
x^{\mu}.
$

The global coordinates on
$
J^{2}_{n}(S)
$
will be
$
(x^{\mu},A_{\lambda},A_{\lambda;\mu},A_{\lambda;\mu\nu}).
$

We particularize all the relevant expressions from the preceding section. The
Euler-Lagrange form is:

\be
T = {\cal T}^{\lambda}~dA_{\lambda} \wedge dx^{1} \wedge ...\wedge dx^{n}.
\ee

To use the main theorem from Section 3, we first define the hyper-Jacobians
(see (\ref{minors}):

\be
\psi_{\rho_{k+1},...,\rho_{n}}^{\mu_{1},...,\mu_{k};\nu_{1},...,\nu_{k}}
\equiv {1\over (n-k)!}
\varepsilon^{\mu_{1},...,\mu_{n}} \varepsilon^{\nu_{1},...,\nu_{n}}
\prod_{i=k+1}^{n} A_{\rho_{i};\mu_{i}\nu_{i}}~~~(k = 0,...,n)
\label{em-Jacobi}
\ee
and the content of the theorem is that we have:

\be
{\cal T}^{\lambda} = \sum_{k=0}^{n} {1\over (k!)^{2}}
t^{\lambda,\rho_{k+1},...,\rho_{n}}_{\mu_{1},...,\mu_{k};\nu_{1},...,\nu_{k}}
\psi_{\rho_{k+1},...,\rho_{n}}^{\mu_{1},...,\mu_{k};\nu_{1},...,\nu_{k}}.
\label{em-polyn}
\ee

Here we suppose that the functions
$
t^{...}_{...}
$
are independent of the second-order derivatives
$
A_{\rho;\mu\nu}
$:

\be
\partial^{\lambda;\zeta\omega}
t^{\rho_{k},...,\rho_{n}}_{\mu_{1},...,\mu_{k};\nu_{1},...,\nu_{k}} = 0
~~~(k = 0,...,n)
\ee
where we have defined:

\be
\partial^{\lambda;\zeta\omega} \equiv
{\partial \over \partial A_{\lambda;\zeta\omega}} \times \cases{
1, & for $\zeta = \omega$ \cr 1/2, & for $\zeta \not= \omega$ \cr},
\ee
have the symmetry properties:

\be
t^{\rho_{P(k)},...,\rho_{P(n)}}_{\mu_{Q(1)},...,\mu_{Q(k)};
\nu_{R(1)},...,\nu_{R(k)}} = (-1)^{|Q|+|R|}
t^{\rho_{k},...,\rho_{n}}_{\mu_{1},...,\mu_{k};\nu_{1},...,\nu_{k}}  
\quad (k = 0,...,n),
\label{antisym-em}
\ee
where
$P$ 
is a permutation of the numbers
$k,...,n$
and 
$Q, R$
are permutations of the numbers 
$1,...,k$,

\be
t^{\rho_{k+1},...,\rho_{n}}_{\nu_{1},...,\nu_{k};\mu_{1},...,\mu_{k}} =
t^{\rho_{k+1},...,\rho_{n}}_{\mu_{1},...,\mu_{k};\nu_{1},...,\nu_{k}} 
\quad (k = 0,...,n)
\label{sym-em}
\ee
and verify the identities:

\be
\sum_{i=0}^{k} (-1)^{i} 
t^{\rho_{k},...,\rho_{n}}_{\mu_{1},...,\mu_{k-1},\nu_{i};
\nu_{0},...,\hat{\nu_{i}},...,\nu_{k}} = 0.
\quad (k = 1,...,n).
\label{identities-em}.
\ee

4.2 We now impose the gauge invariance of the theory.
The gauge group, denoted by
$
Gau(U(1))
$
consists of smooth maps
$
g: M \rightarrow U(1)
$
with pointwise multiplication as composition law. We will consider only
infinitesimal transformation
$
g(x) \cong {\bf1} + i\xi(x)
$
where
$
\xi: M \rightarrow \R
$
is a smooth function. The gauge group acts infinitesimally  as follows:

\be
\phi_{\xi}(x^{\mu},A_{\lambda}) =
(x^{\mu},A_{\lambda}+(\partial_{\lambda} \xi)(x))
\label{Abelian-gauge}
\ee

and we impose the condition of gauge invariance as follows:

\be
(j^{2}\phi_{\xi})^{*} T = T
\label{Abelian-inv}
\ee
(see subsection 2.6).

From (\ref{Abelian-gauge}) we easily derive:

\be
\begin{array}{c}
j^{2}\phi_{\xi} (x^{\mu},A_{\lambda},A_{\lambda;\mu},A_{\lambda;\mu\nu}) =
\nonumber \\
(x^{\mu},A_{\lambda}+(\partial_{\lambda} \xi)(x),
A_{\lambda;\mu}+(\partial_{\lambda}\partial_{\mu} \xi)(x),
A_{\lambda;\mu\nu}+ (\partial_{\lambda}\partial_{\mu}\partial_{\nu} \xi)(x))
\end{array}
\ee
and the invariance condition (\ref{Abelian-inv}) we get, equivalently, the
following set of relations:

\be
\partial^{\mu}~T^{\lambda} = 0
\ee

\be
(\partial^{\zeta;\mu} + \partial^{\mu;\zeta} )~T^{\lambda} = 0
\ee

\be
\sum_{\zeta,\mu,\nu} \partial^{\zeta;\mu\nu}~T^{\lambda} = 0
\ee
where

\be
\partial^{\zeta} \equiv {\partial \over \partial A_{\zeta}}
\ee
and
\be
\partial^{\zeta;\mu} \equiv {\partial \over \partial A_{\zeta;\mu}}.
\ee

Now we insert in the preceding equations the expression of
$
T^{\lambda}
$
given by (\ref{em-polyn}) and we get equivalently:

\be
\partial^{\mu}~t^{\rho_{0},...,\rho_{k}} = 0~~~(k = 0,...,n)
\label{inv1}
\ee

\be
(\partial^{\zeta;\mu} + \partial^{\mu;\zeta} )~
t^{\rho_{0},...,\rho_{k}} = 0~~~(k = 0,...,n)
\label{inv2}
\ee

\be
\sum_{\zeta,\mu,\nu}
\left[ b^{*(\mu)} c^{*(\nu)} + b^{*(\nu)} c^{*(\mu)} \right]~
t^{\zeta,\rho_{1},...,\rho_{k}} = 0~~~(k = 1,...,n)
\label{inv3}
\ee
where we again we tensor notations. The equation (\ref{inv3}) can be used to
make some essential simplification. First we easily establish
that we (\ref{identities-em}) is equivalent to

\be
{\cal C} t^{\rho_{0},...,\rho_{p}} = 0, \quad 
{\cal C}^{*} t^{rho_{0},...,\rho_{p}} = 0 \quad (p = 0,...,n).
\ee

Then, we can apply lemma \ref{perm} and obtain that 
$
t^{\rho_{0},...,\rho_{k}}
$
has the following form:

\be
t^{\rho_{0},...,\rho_{k}} = b^{*(\rho_{0})}~C^{\rho_{1},...,\rho_{k}} +
c^{*(\rho_{0})}~SC^{\rho_{1},...,\rho_{k}}~~~(k = 1,...,n)
\label{structure-t}
\ee
with
$
C^{\rho_{1},...,\rho_{k}} \in {\cal H}^{*}_{0,n-k-1,n-k}
$
and obvious symmetry properties. If
$
k >1
$
we use the symmetry property in
$
\rho_{0},...,\rho_{k}
$
to obtain 

\be
t^{\rho_{0},...,\rho_{k}} = b^{*(\rho_{1})}
\quad C^{\rho_{0},\rho_{2},...,\rho_{k}} +
c^{*(\rho_{0})} \quad SC^{\rho_{0},\rho_{2},...,\rho_{k}} \qquad
(k = 1,...,n)
\ee

The last two equations can be combined to obtain that 
$
t^{\rho_{0},...,\rho_{k}}
$
has in fact the following form:

\be
t^{\rho_{0},...,\rho_{k}} = 
b^{*(\rho_{0})}  b^{*(\rho_{1})}  C^{\rho_{2},...,\rho_{k}} +
b^{*(\rho_{0})}  c^{*(\rho_{1})}  D^{\rho_{2},...,\rho_{k}} +
c^{*(\rho_{0})}  c^{*(\rho_{1})}  \tilde{C}^{\rho_{2},...,\rho_{k}} +
c^{*(\rho_{0})}  b^{*(\rho_{1})}  \tilde{D}^{\rho_{2},...,\rho_{k}}.
\ee

Keeping only the part which is symmetric in
$
\rho_{0}
$
and
$
\rho_{1}
$
we obtain a simpler expression:

\be
t^{\rho_{0},...,\rho_{k}} = 
[b^{*(\rho_{0})}  c^{*(\rho_{1})} + 
b^{*(\rho_{1})} c^{*(\rho_{0})}] 
\quad (D^{\rho_{2},...,\rho_{k}} + \tilde{D}^{\rho_{2},...,\rho_{k}}).
\ee

Finally, if 
$
p \geq 2
$
we have from the symmetry properties and (\ref{structure-t}):

\be
t^{\rho_{0},...,\rho_{k}} = b^{*(\rho_{2})}
\quad C^{\rho_{0},\hat{\rho_{2}},...,\rho_{k}} +
c^{*(\rho_{0})} \quad SC^{\rho_{0},\hat{\rho_{2}},...,\rho_{k}} \qquad
(k = 1,...,n)
\ee
which can be combined with the preceding equation to obtain:

\be
t^{\rho_{0},...,\rho_{k}} = 
[b^{*(\rho_{0})} c^{*(\rho_{1})} + 
b^{*(\rho_{1})} c^{*(\rho_{0})}] 
\left[ b^{*(\rho_{2})} C^{\rho_{0},\hat{\rho_{2}},...,\rho_{k}} +
c^{*(\rho_{0})} SC^{\rho_{0},\hat{\rho_{2}},...,\rho_{k}}
\right] \qquad (k = 1,...,n).
\ee

Taking the part which is symmetric in
$
\rho_{0}
$
and
$
\rho_{2}
$
of this relation we finally obtain:

\be
t^{\rho_{0},...,\rho_{k}} = 0 \qquad (k > 2).
\ee

So, it follows that for the electromagnetic field the Euler-Lagrange
expressions are at most linear in the second order derivatives i.e. we are in
the particular case studied in Section 3.4. If we apply the reduction of
the order theorem we can conclude that these Euler-Lagrange expressions
can be obtained from a first-order Lagrangian. But in this case, we are
back in the framework studied in \cite{G2} and we can conclude that we have:

\begin{thm}

Every second-order Euler-Lagrange expression for the pure electromagnetic
field  follows from a first-order Lagrangian which can be taken to be the
sum of a part depending only on the field strength and a second part which
is the Chern-Simons Lagrangian.
\end{thm}

We will provide below an independent proof of this fact using Fock space
techniques.

4.3 It is convenient to work with new functions as in section 3.4, namely:

\be
\tilde{t}^{\lambda} \equiv {1 \over (n!)^{2}}
\varepsilon^{\mu_{1},...,\mu_{n}} \varepsilon^{\nu_{1},...,\nu_{n}}
t^{\lambda}_{\mu_{1},...,\mu_{n};\nu_{1},...,\nu_{n}}
\ee

and

\be
\tilde{t}^{\lambda,\zeta;\mu_{1}\nu_{1}} \equiv {1 \over ((n-1)!)^{2}}
\varepsilon^{\mu_{1},...,\mu_{n}} \varepsilon^{\nu_{1},...,\nu_{n}}
t^{\lambda,\zeta}_{\mu_{2},...,\mu_{n};\nu_{2},...,\nu_{n}}.
\ee

We note the symmetry property:

\be
\tilde{t}^{\lambda,\zeta;\mu\nu} = \tilde{t}^{\zeta\lambda;\mu\nu} =
\tilde{t}^{\lambda,\zeta;\nu\mu}
\ee
and the fact that one can rewrite the Euler-Lagrange expressions as follows:

\be
{\cal T}^{\lambda} = \tilde{t}^{\lambda\zeta;\mu\nu}~A_{\zeta;\mu\nu} +
\tilde{t}^{\lambda}.
\label{EL-sec}
\ee

Then the invariance conditions (\ref{inv1}) and (\ref{inv2}) tell us that the
functions
$
\tilde{t}^{\lambda\zeta;\mu\nu}
$
and
$
\tilde{t}^{\lambda}
$
depend only of
$
x^{\mu}
$
and of the field strength

\be
F_{\mu\nu} \equiv A_{\mu;\nu} - A_{\nu;\mu}
\label{F}
\ee
and (\ref{inv3}) becomes a purely algebraic condition:

\be
\sum_{\lambda,\mu,\nu}~\tilde{t}^{\lambda,\zeta;\mu,\nu} = 0.
\label{inv3'}
\ee

According to the order reduction theorem (more precisely the corollary after
it) one can find a first-order Lagrangian such that

\be
{\cal T}^{\lambda} = {\cal E}^{\lambda}({\cal L}).
\ee

In particular we have

\be
\tilde{t}^{\lambda\zeta;\mu\nu} = -{1\over 2}
(\partial^{\lambda;\mu}\partial^{\zeta;\nu} +
\partial^{\lambda;\nu}\partial^{\zeta;\mu}){\cal L}
\label{EfromL}
\ee
for a function
${\cal L}$
independent of the second-order derivatives. The point is to show that this
function (which is not uniquely determined by the previous equation) can be
chosen to have the same dependence as
$
\tilde{t}^{\lambda\zeta;\mu\nu}
$
i.e. only of the space-time
$
x^{\mu}
$
and of the field strength $F$.

One can derive rather easily from (\ref{EfromL}) that the Lagrangian function
${\cal L}$
verifies the following equations:

\be
(\partial^{\lambda;\mu}\partial^{\zeta;\nu} +
\partial^{\lambda;\nu}\partial^{\zeta;\mu})~\partial^{\zeta}{\cal L} = 0
\label{invI}
\ee

\be
(\partial^{\lambda;\mu}\partial^{\zeta;\nu} +
\partial^{\lambda;\nu}\partial^{\zeta;\mu})~
(\partial^{\zeta;\rho} + \partial^{\rho;\zeta}) {\cal L} = 0.
\label{invII}
\ee

We need now the expression of the kernel of the operator
$
\partial^{\lambda;\mu}\partial^{\zeta;\nu} +
\partial^{\lambda;\nu}\partial^{\zeta;\mu}.
$

This is has been obtained in corollary 1 and is in our case:

\be
{\cal L} = \sum_{k=0}^{n} {1\over k!}
C^{\lambda_{1},...,\lambda_{k};\mu_{1},...,\mu_{k}}
\prod_{i=0}^{k} A_{\lambda_{i};\mu_{i}}
\label{trivial}
\ee
where the functions
$
C^{\lambda_{1},...,\lambda_{k};\mu_{1},...,\mu_{k}}
$
do not depend on the second order-derivatives and are completely
antisymmetric in the indices
$
\lambda_{1},...,\lambda_{k}
$
and in the indices
$
\mu_{1},...,\mu_{k}.
$

Applying this lemma to the relation (\ref{invII}) we obtain that

\be
(\partial^{\lambda;\mu}\partial^{\zeta;\nu} +
\partial^{\lambda;\nu}\partial^{\zeta;\mu}) {\cal L} =
\sum_{k=0}^{n} {1\over k!}
C^{\mu,\nu;\zeta_{1},...,\zeta_{k};\rho_{1},...,\rho_{k}}
\prod_{i=0}^{k} A_{\zeta_{i};\rho_{i}}
\ee
where, beside the symmetry properties from the lemma above the coefficients
$
C^{...}
$
are also symmetric in
$\mu$
and
$\nu.$

One gets from the preceding equation an integrability condition, namely:

\be
C^{\rho\sigma;\lambda,\mu_{1},...,\mu_{k};\zeta,\nu_{1},...,\nu_{k}} =
(\rho,\sigma \leftrightarrow \lambda,\zeta)~~~(k = 0,...,n-1).
\ee

To analyze this system we need 

\begin{lemma}

Let
$
C^{\mu\nu} \in {\cal H}_{0,k,k} \quad (1 < k \leq n)
$
verifying

\be
C^{\mu\nu} = C^{\nu\mu}
\ee
and the system:

\be
[b^{(\lambda)} c^{(\zeta)} + b^{(\zeta)} c^{(\lambda)}] C^{\rho,\sigma} =
(\rho,\sigma \leftrightarrow \lambda,\zeta).
\label{CC}
\ee

Then there exists 
$
C \in {\cal H}_{0,k+1,k+1}
$
such that:

\be
C^{\mu\nu} = (b^{(\mu)} c^{(\nu)} + b^{(\nu)} c^{(\mu)}) C.
\ee
\end{lemma}

{\bf Proof:}
(i) We define as usual

\be
{\cal C} \equiv c^{*}_{(\mu)} b^{(\mu)}, \quad
{\cal A} \equiv {\cal C}^{*} {\cal C}
\ee

and apply the operator 
$
b^{*}_{(\mu)} c^{*}_{(\nu)} 
$
to the equation (\ref{CC}). We obtain easily:

\be
\left[ \left( k(k+1) -2\right) {\bf 1} - {\cal A} \right] C^{\mu\nu} =
\left[ b^{(\mu)} c^{(\nu)} + b^{(\nu)} c^{(\mu)} \right] T -
b^{(\mu)} C^{\nu} - c^{(\mu)} D^{\nu} -
b^{(\nu)} C^{\mu} - c^{(\nu)} D^{\mu}
\label{TCD}
\ee
where
$
C^{\mu} \equiv b^{*}_{(\rho)} C^{\rho\mu}, \quad 
D^{\mu} \equiv c^{*}_{(\rho)} C^{\rho\mu}, \quad
$
and
$
T \equiv b^{*}_{(\rho)} c^{*}_{(\sigma)} C^{\rho\sigma}.
$

(ii) We will prove here that 
$
\lambda = k(k+1) -2
$
cannot be an eigenvalue of the operator
$
{\cal A}.
$
First one proves by induction that if

\be
{\cal A} t = \lambda t
\label{eigen}
\ee
then

\be
{\cal A} {\cal C}^{s} t = [\lambda - (s+1)s] {\cal C}^{s} t 
\quad s = 0,1,...k.
\label{eigenS}
\ee

We prove now that
\be \label{Ck} 
{\cal C}^{k} t = 0.
\ee

For
$
2k  > n
$
this is trivial. In the opposite case we take
$
s = k
$
in (\ref{eigenS}) and we have:

\be
{\cal A} {\cal C}^{k} t = - 2 {\cal C}^{k} t.
\ee

But
$
{\cal A } {\cal C}^{k} t = {\cal C}^{*} {\cal C}^{k+1} t = 0
$
so we obtain again (\ref{Ck}).

We apply to the equation (\ref{Ck}) the operator
$
{\cal C}^{*}
$
and obtain
$
\lambda_{k-1} {\cal C}^{k-1} t = 0
$
i.e.
$
{\cal C}^{k-1} t = 0
$
because
$
\lambda_{k-1} \not= 0.
$
Iterating the procedure one gets finally
$
t = 0
$
i.e. a contradiction. This shows that
$
k(k + 1) - 2
$
cannot be an eigenvalue of the operator
$A$.

(iii)
One can apply again to (\ref{TCD}) the operators
$
b^{*}_{(\mu)}
$
and
$
c^{*}_{(\nu)}
$
respectively and after some rather long computations we obtain relations of
the type:

\be
A_{c} C^{\mu} = b^{(\mu)} \cdots + c^{(\mu)} \cdots
\ee
and

\be
A_{d} D^{\mu} = b^{(\mu)} \cdots + c^{(\mu)} \cdots
\ee
where the operators
$
A_{b}
$
and
$
A_{d}
$
are of the type 
$
{\cal A}^{2} + \alpha^{2} {\bf 1}
$
with
$
\alpha \in \R^{*}.
$
So
$
A_{c}
$
and
$
A_{d}
$
are invertible. Using the finite dimensional calculus one can obtain that
the tensors
$
C^{\mu}
$
and
$
D^{\mu}
$
are expressions of the type
$
b^{(\mu)} \cdots + c^{(\mu)} \cdots.
$

Plugging this result into (\ref{TCD}) and keeping only the part which is
symmetric in the indices
$\mu$
and
$\nu$
we obtain a relation of the type

\be
\left[ \left(k (k+1) -2\right) {\bf 1} - {\cal A} \right] C^{\mu\nu} =
\left[ b^{(\mu)} c^{(\nu)} + b^{(\nu)} c^{(\mu)} \right] T_{0}.
\ee

Using (ii) and the finite dimensional calculus one easily
obtains that
$
C^{\mu\nu}
$
has the expression from the statement.
\qed

Using index notations we have:

\be
C^{\rho\sigma;\mu_{1},...,\mu_{k};\nu_{1},...,\nu_{k}} =
C^{\rho,\mu_{1},...,\mu_{k};\sigma,\nu_{1},...,\nu_{k}} +
C^{\sigma,\mu_{1},...,\mu_{k};\rho,\nu_{1},...,\nu_{k}}
\ee
for some
$
C^{k} \in {\cal H}^{*}_{0,k-1,k-1}, \quad k \geq 2.
$

We analyze separately the case
$
k = 1
$
and easily obtain that

\be
C^{\mu\nu;\rho;\sigma} = P^{\mu\nu;\rho\sigma} + Q^{\mu\nu;\rho\sigma}
\ee
where we have the following symmetry properties:

\be
P^{\mu\nu;\rho\sigma} = P^{\rho\sigma;\mu\nu} =
P^{\nu\mu;\rho\sigma} = P^{\mu\nu;\sigma\rho}
\ee
and

\be
Q^{\mu\nu;\rho\sigma} = Q^{\nu\mu;\rho\sigma} = - Q^{\mu\nu;\sigma\rho}.
\ee
 
Now, using the corollary from the preceding section one can easily show that
the redefinition of the Lagrangian:

\be
{\cal L} \rightarrow {\cal L} + {1\over 2} C^{\mu\nu}
A_{\nu\{\mu\}} \sum_{k=2}^{n} {1\over k!}
C^{\lambda_{1},...,\lambda_{k};\mu_{1},...,\mu_{k}}
\prod_{i=0}^{k} A_{\lambda_{i};\mu_{i}}
\ee
does not modify the starting point (\ref{EfromL}) and neither (\ref{invII}).
Moreover, in this way the new Lagrangian will verify the equation:

\be
(\partial^{\rho;\sigma} + \partial^{\sigma;\rho}) {\cal L} = 
{1\over 8} P^{\rho\sigma;\mu\nu} ( A_{\nu\{\mu\}} + A_{\mu\{\nu\}}) - 
{1\over 8} Q^{\rho\sigma;\mu\nu} F_{\mu\nu}.
\ee

This equation can be integrated and afterwards use can be made
of the invariance condition (\ref{inv3'}). As a result we get that
$
{\cal L}
$
can be found out such that beside (\ref{EfromL}) and
(\ref{invII}) also verifies:

\be
(\partial^{\rho;\sigma} + \partial^{\sigma;\rho}) {\cal L} = 0.
\ee

Using again the corollary
from the preceding section one can show using analogous tricks that the
Lagrangian can be redefined as above such that it will not depend on
$
A_{\lambda}.
$

We have proved our assertion, namely that we can find a Lagrangian
${\cal L}_{0}$
dependent only on
$
x^{\mu}
$
and
$F$
and such that (\ref{EfromL}) is true. Moreover, one can show easily that the
algebraic condition (\ref{inv3'}) is now an identity. Because the Lagrangian
$
{\cal L}_{0}
$
is gauge invariant, it easily follows that the associated Euler-Lagrange
expressions are gauge invariant:

\be
{\cal E}^{\lambda}({\cal L}_{0}) \circ j^{2}\phi_{\xi} =
{\cal E}^{\lambda}({\cal L}_{0}).
\ee

4.4 Let us define now the expressions:

\be
{\cal T}^{\lambda}_{CS} \equiv {\cal T}^{\lambda} -
{\cal E}^{\lambda}({\cal L}_{0}).
\ee

It is clear that
$
{\cal T}^{\lambda}_{CS}
$
will be gauge invariant Euler-Lagrange expressions. Moreover they will not
depend on the second-order derivatives. (Compare relation (\ref{EfromL}) with
the expression (\ref{EL-sec})). So, they will depend only on
$
x^{\mu}
$
and
$F$.
It is not very complicated to show that the ADK equations reduce in this case
to the following relations:

\be
{\partial {\cal T}^{\lambda}_{CS} \over \partial F_{\zeta\rho}} +
{\partial {\cal T}^{\zeta}_{CS} \over \partial F_{\lambda\rho}}  = 0
\label{CS1}
\ee
and

\be
{\partial \over \partial x^{\rho}} \left(
{\partial {\cal T}^{\lambda}_{CS} \over \partial F_{\zeta\rho}} -
{\partial {\cal T}^{\zeta}_{CS} \over \partial F_{\lambda\rho}} \right) = 0.
\label{CS2}
\ee

Now is rather elementary to obtain from (\ref{CS1}) a polynomial structure
for
$
{\cal T}^{\lambda}_{CS};
$
we get

\be
{\cal T}^{\lambda}_{CS} = \sum_{k=0}^{n} {1 \over k!}
C^{\lambda,\mu_{1},...,\mu_{k},\nu_{1},...,\nu_{k}}
\prod_{i=0}^{k} F_{\mu_{i}\nu_{i}}
\label{Chern}
\ee
where the expressions
$
C^{...}
$
are completely antisymmetric in all indices and can depend only on
$
x^{\mu}.
$
Moreover, from (\ref{CS2}) the dependence on
$
x^{\mu}
$
is cancelled, i.e. the expressions
$
C^{...}
$
from (\ref{Chern}) are in fact constants. A corresponding Lagrangian can be
obtained applying Tonti formula; one get the usual expression:

\be
{\cal L}_{CS}= A^{\lambda} \sum_{k=0}^{n} {1\over (k+1)!}
C^{\lambda,\mu_{1},...,\mu_{k},\nu_{1},...,\nu_{k}} \prod_{i=1}^{k}
F_{\nu_{i}\mu_{i}}.
\ee

\begin{rem}

If one imposes Poincar\'e invariance also, one can prove the following facts.
First, as in subsection 4.4 one can redefine he Lagrangian
$
{\cal L}_{0}
$
such that it is in fact Poincar\'e invariant. This will mean that it will be
only
$F$-dependent
and also a Lorentz scalar. Some cohomology arguments of the type used in
\cite{G2} must be used. Then it follows that the Chern-Simons part is also
Poincar\'e invariant. This in turn means that the constants
$
C^{...}
$
from (\ref{Chern}) will be the components of a Lorentz invariant tensor. Taking
into account the complete antisymmetry on obtains the usual result: the
Chern-Simons Lagrangian can be constructed only in odd-dimensional spaces;
if
$
n = 2m +1
$
then in (\ref{Chern}) only the term corresponding to
$
k = m
$
survives and the tensor
$
C^{..}
$
must be replaced by a constant times the corresponding completely
antisymmetric tensor.
\end{rem}

\begin{rem}
Similar results about the order reduction of the variational problem can be
proved for other interesting theories as pure Yang-Mills theory or pure
gravitational theory. More precisely, one can again discover from the
corresponding invariance conditions (gauge invariance and reparametrization
invariance respectively) that the Euler-Lagrange expressions are at most
linear in the second-order derivatives. So, one can obtain these expressions
from first-order Lagrangians. But in this case one can use for the
corresponding analysis the formalism from \cite{G2}. An direct analysis on the
lines followed above is also possible. On the contrary, in the case of
extended objects (as strings, see \cite{G2}, part 9) such a reduction of the
order does not exists, so one can have more general second-order
Euler-Lagrange expressions that those following from the Nambu-Goto
Lagrangian.
\end{rem}

\section{Conclusions}

We have given a complete analysis of the most general form of a
second-order partial differential equation of the Lagrangian type
starting from the Anderson-Duchamp-Krupka equations. The next
logical step would be to try to extend this analysis to the case
of Grassmann variables. In this way one would be able to study
Lagrangian systems with BRST-type symmetries.

\vskip 1cm

{\bf Acknowledgments:} The author thanks professor Krupka for many
interesting discussions on the subject of variational calculus. The visit
at Opava University was made possible by the grant ``Global Variational
Functionals of Mathematical Physics" of the Ministry of Education and Youth,
No. 871/1995

\end{document}